\documentclass[12pt,a4]{article}
\usepackage{amsmath}
\usepackage{bm}
\usepackage{amsfonts}
\usepackage{amssymb}
\usepackage{graphics}
\usepackage[normal]{caption2}
\usepackage{subfigure}
\usepackage{rotating}
\usepackage{citesort}
\def\be{\begin{equation}}
\def\ee{\end{equation}}
\def\ba{\begin{array}}
\def\ea{\end{array}}
\def\bea{\begin{eqnarray}}
\def\eea{\end{eqnarray}}

\begin{document}
\baselineskip 20pt \setlength\tabcolsep{2.5mm}
\renewcommand\arraystretch{1.5}
\setlength{\abovecaptionskip}{0.1cm}
\setlength{\belowcaptionskip}{0.5cm}
\pagestyle{empty}
\newpage
\pagestyle{plain} \setcounter{page}{1} \setcounter{lofdepth}{2}
\begin{center} {\large\bf System size effects in the N/Z dependence of balance energy for isotopic series}\\
\vspace*{0.4cm}

{\bf Sakshi Gautam$^a$} and {\bf Aman D. Sood$^{b}$}\footnote{Email:~amandsood@gmail.com}\\
$^a${\it  Department of Physics, Panjab University, Chandigarh
-160 014, India.\\}$^{b}$ {\it  SUBATECH, Laboratoire de Physique Subatomique et des
Technologies Associ\'{e}es, Universit\'{e} de Nantes - IN2P3/CNRS
- EMN 4 rue Alfred Kastler, F-44072 Nantes, France.\\}
\end{center}
We study the system size effects in the N/Z dependence of balance energy for the isotopic series. We find
drastic effect of symmetry energy on the N/Z dependence of
E$_{bal}$ throughout the mass range. We also find that the N/Z
dependence of E$_{bal}$ for isotopic series of lighter system is
slightly more sensitive to symmetry energy as compared to that of
heavier systems. We also study the mass dependence of E$_{bal}$
for the N/Z range from 1.0-2.0. We find that the mass dependence
of E$_{bal}$ varies with the N/Z ratio.


\newpage
\baselineskip 20pt
\section{Introduction}
 The investigation of
the system size effects in various phenomena of heavy-ion
collisions has attracted a lot of attention. The system size
dependences have been reported in various phenomena like
fusion-fission, particle production, multifragmentation,
collective flow (of nucleons/fragments) as well as its
disappearance, density, temperature and so on
\cite{puri,huang,jian06,stur01,sing,mag1,balirep,sood4,sood2,sood3,sood1,ogli,pan,zhang,krof}.
For instance, in Ref. \cite{stur01} the power law scaling
($\varpropto$ A$^{\tau}$) of pion/kaon production with the size of
the system has been reported. Similar power law behavior for the
system size dependence has been reported for the multiplicity of
various types of fragments also \cite{sing}. The collective
transverse in-plane flow has also been investigated extensively
during the past three decades and has been found to depend
strongly on the combined mass of the system \cite{ogli} in
addition to the incident energy \cite{pan,zhang} as well as
colliding geometry \cite{zhang}. The energy dependence of
collective transverse in-plane flow has led us to its
disappearance. The energy at which flow disappears has been termed
as the balance energy (E$_{bal}$) or the energy of vanishing flow
(EVF) \cite{krof}. E$_{bal}$ has been found to depend strongly on
the combined mass of the system \cite{mag1,balirep}. Similarly the power
law mass dependences have also been reported for density,
temperature, and participant-spectator matter \cite{sood4}.
\par
 With the advent of radioactive ion beams \cite{rib1,rib2,rib3} the role of isospin degree of freedom on dynamics
 of heavy-ion collisions has been studied for the past decade. These studies are helpful to
 extract information about the asymmetric nuclear matter. The isospin effects have been explained in literature as the competition
 among various reaction mechanisms, such as nn collisions, symmetry energy, surface properties and Coulomb
 force. The relative importance among these reaction mechanisms
 is not yet clear \cite{li}. Therefore, to explain the relative contribution among these reaction mechanisms, Gautam \emph{at al}. have studied
E$_{bal}$ as a function of combined mass of the system
\cite{gaum10} as well as colliding geometry for isobaric pairs
\cite{gaum210}. There we find that throughout the mass range and colliding geometry, the systems with higher N/Z has larger E$_{bal}$ as compared to
 the systems with lower N/Z. This was found to be due to the dominance of Coulomb repulsion over symmetry energy in isospin effects throughout
 the mass range and colliding geometry. Therefore,
 to take out the role of dominant Coulomb Sod has studied the E$_{bal}$ for a series of isotopes of Ca with N/Z varying from 1.0 to 2.0 \cite{gaum310}.
 Sood has found that although for an individual isotope E$_{bal}$ is sensitive to symmetry energy as well as its density dependence, isospin-dependent
 cross section, equation of state (EOS) of symmetric nuclear matter simultaneously, however the N/Z dependence of E$_{bal}$ showed sensitivity
 not only
 to symmetry energy but more
 importantly to the density dependence of symmetry energy as well.
In the present paper, our aim is at least two fold.\\
(1) To study N/Z dependence of E$_{bal}$ for isotopic series throughout the mass range and also to explore the effect of symmetry energy. \\
(2) To study mass dependence of E$_{bal}$ for various N/Z ratios
 covering pure symmetric systems to highly neutron-rich ones.
\par For the present study we use isospin-dependent quantum molecular dynamics (IQMD) model
\cite{hart98}.

\section{The Model}
 The IQMD model is an extension of the QMD model \cite{aichqmd,lehmann}, which treats different charge states of
nucleons, deltas and pions explicitly, as inherited from the
Vlasov-Uehling-Uhlenbeck (VUU) model. The IQMD model has been used
successfully for the analysis of a large number of observables
from low to relativistic energies. The isospin degree of freedom
enters into the calculations via symmetry potential, cross
sections and Coulomb interaction.
 \par
 In this model, baryons are represented by Gaussian-shaped density distributions
  \begin {eqnarray}
  f_{i}(\vec{r},\vec{p},t) =
  \frac{1}{\pi^{2}\hbar^{2}}\exp(-[\vec{r}-\vec{r_{i}}(t)]^{2}\frac{1}{2L})
   \times \exp(-[\vec{p}- \vec{p_{i}}(t)]^{2}\frac{2L}{\hbar^{2}})
 \end {eqnarray}
 Nucleons are initialized in a sphere with radius R = 1.12 A$^{1/3}$ fm, in accordance with the liquid-drop model.
 Each nucleon occupies a volume of \emph{h$^{3}$}, so that phase space is uniformly filled.
 The initial momenta are randomly chosen between 0 and Fermi momentum ($\vec{p}$$_{F}$).
 The nucleons of the target and projectile interact by two- and three-body Skyrme forces, Yukawa potential, and Coulomb interactions. In addition to the use of explicit charge states of all baryons and mesons, a symmetry potential between protons and neutrons
 corresponding to the Bethe-Weizsacker mass formula has been included.
 The hadrons propagate using the Hamilton equations of motion:
 \begin {eqnarray}
 \frac{d\vec{{r_{i}}}}{dt} = \frac{d\langle H
 \rangle}{d\vec{p_{i}}};\frac{d\vec{p_{i}}}{dt} = - \frac{d\langle H
  \rangle}{d\vec{r_{i}}}
 \end {eqnarray}
  with
  \begin {eqnarray}
  \nonumber\langle H\rangle =\langle T\rangle
+ \langle V \rangle \\
                  \nonumber= \sum_{i}\frac{p^{2}_{i}}{2m_{i}} +
                   \sum_{i}\sum_{j>i}\int
                   f_{i}(\vec{r},\vec{p},t)V^{ij}
                   (\vec{r}~',\vec{r}) \\\times
                   f_{j}(\vec{r}~',\vec{p}~',t)
                   d\vec{r}~ d\vec{r}~'~ d\vec{p}~ d\vec{p}~'.
 \end {eqnarray}
 The baryon potential\emph{ V$^{ij}$}, in the above relation, reads as
 \begin {eqnarray}
  \nonumber V^{ij}(\vec{r}~'-\vec{r}) = V^{ij}_{Skyrme} + V^{ij}_{Yukawa} +
  V^{ij}_{Coul} + V^{ij}_{sym} \\
    \nonumber=[t_{1}\delta(\vec{r}~'-\vec{r})+t_{2}\delta(\vec{r}~'-\vec{r})\rho^{\gamma-1}(\frac{\vec{r}~'+\vec{r}}{2})]\\
   \nonumber ~+t_{3}\frac{\exp(|(\vec{r}~'-\vec{r})|/\mu)}{(|(\vec{r}~'-\vec{r})|/\mu)}+
    \frac{Z_{i}Z_{j}e^{2}}{|(\vec{r}~'-\vec{r})|}\\
    +t_{4}\frac{1}{\varrho_{0}}T_{3i}T_{3j}\delta(\vec{r_{i}}~'-\vec{r_{j}}).
 \end {eqnarray}
Here t$_{4}$ = 4C with C = 32 MeV and \emph{Z$_{i}$} and
\emph{Z$_{j}$} denote the charges of the \emph{ith} and \emph{jth}
baryon, and \emph{T$_{3i}$} and
 \emph{T$_{3j}$}
 are their respective \emph{T$_{3}$} components (i.e. $1/2$ for protons and $-1/2$ for neutrons).
The parameters\emph{ $\mu$} and \emph{t$_{1}$,....,t$_{4}$} are
adjusted to the real part of the nucleonic optical potential.
 For the density dependence of  the nucleon optical potential, standard Skyrme-type parametrization is employed.
We use a soft
equation of state along with the standard isospin- and
energy-dependent cross section reduced by
  20$\%$, i.e. $\sigma$ = 0.8 $\sigma_{nn}^{free}$. For the density dependence of symmetry energy,
 we use the form F$_{1}(u) \propto u$ (where \emph{u} =
$\frac{\rho}{\rho_{0}}$) which is obtained by changing the density dependence of potential part
 of symmetry energy (last term in equation 4).
  F$_{2}$ represents
calculations without symmetry energy. The symmetry energy is switched off by making the potential part of symmetry energy zero.
In a recent study, Gautam \emph{et al}. \cite{gaum210}
  has confronted the theoretical calculations of IQMD with the data of $^{58}Ni+^{58}Ni$ and $^{58}Fe+^{58}Fe$ \cite{pak97}.
  The results with the soft EOS (along with the
momentum-dependent interactions) and above choice of cross section are in good agreement with the data at all colliding geometries.
The details about the elastic and inelastic cross sections for
proton-proton and proton-neutron collisions can be found in
\cite{hart98,cug}. The cross sections for neutron-neutron
collisions are assumed to be equal to the proton-proton cross
sections. Two particles collide if their minimum distance\emph{ d}
fulfills
\begin {equation}
 d \leq d_{0} = \sqrt{\frac{\sigma_{tot}}{\pi}},   \sigma_{tot} =
 \sigma(\sqrt{s}, type),
\end {equation}
where 'type' denotes the ingoing collision partners (N-N....).
Explicit Pauli blocking is also included; i.e. Pauli blocking of
the neutrons and protons is treated separately. We assume that
each nucleon occupies a sphere in coordinate and momentum space.
This trick yields the same Pauli blocking ratio as an exact
calculation of the overlap of the Gaussians will yield. We
calculate the fractions P$_{1}$ and P$_{2}$ of final phase space
for each of the two scattering partners that are already occupied
by other nucleons with the same isospin as that of scattered ones.
The collision is blocked with the probability
\begin {equation}
 P_{block} = 1-[1 - min(P_{1},1)][1 - min(P_{2},1)],
\end {equation}
and, correspondingly is allowed with the probability 1 -
P$_{block}$. For a nucleus in its ground state, we obtain an
averaged blocking probability $\langle P_{block}\rangle$ = 0.96.
Whenever an attempted collision is blocked, the scattering
partners maintain the original momenta prior to scattering.

\par

\section{Results and discussions}

We simulate the reactions of Ca+Ca, Ni+Ni, Zr+Zr, Sn+Sn, and Xe+Xe
with N/Z varying from 1.0 to 2.0 in small steps of 0.2. In
particular we simulate the reactions of $^{40}$Ca+$^{40}$Ca
(N/Z=1), $^{44}$Ca+$^{44}$Ca (N/Z=1.2), $^{48}$Ca+$^{48}$Ca
(N/Z=1.4), $^{52}$Ca+$^{52}$Ca (N/Z=1.6), $^{56}$Ca+$^{56}$Ca
(N/Z=1.8), and $^{60}$Ca+$^{60}$Ca (N/Z=2.0); $^{56}$Ni+$^{56}$Ni,
$^{62}$Ni+$^{62}$Ni, $^{68}$Ni+$^{68}$Ni, $^{72}$Ni+$^{72}$Ni, and
$^{78}$Ni+$^{78}$Ni; $^{81}$Zr+$^{81}$Zr, $^{88}$Zr+$^{88}$Zr,
$^{96}$Zr+$^{96}$Zr, $^{104}$Zr+$^{104}$Zr, and
$^{110}$Zr+$^{110}$Zr; $^{100}$Sn+$^{100}$Sn,
$^{112}$Sn+$^{112}$Sn, $^{120}$Sn+$^{120}$Sn,
$^{129}$Sn+$^{129}$Sn, and $^{140}$Sn+$^{140}$Sn; and
$^{110}$Xe+$^{110}$Xe, $^{120}$Xe+$^{120}$Xe,
$^{129}$Xe+$^{129}$Xe, $^{140}$Xe+$^{140}$Xe, and
$^{151}$Xe+$^{151}$Xe at b/b$_{max}$ = 0.2 - 0.4. The reactions are followed till the transverse in-plane
flow saturates. It is worth mentioning here that the saturation time
varies with the mass of the system. It has been shown in Ref. \cite{sood1} that the transverse in-plane
flow in lighter colliding nuclei saturates earlier compared to heavy colliding nuclei. Saturation time is about 100 (200 fm/c) in
lighter (heavy) colliding nuclei in the present energy domain. We use the quantity "\textit{directed transverse
momentum $\langle p_{x}^{dir}\rangle$}" to define the nuclear
transverse in-plane flow, which is defined as
\cite{sood1,hart98,aichqmd,leh}
\begin {equation}
\langle{p_{x}^{dir}}\rangle = \frac{1} {A}\sum_{i=1}^{A}{sign\{
{y(i)}\} p_{x}(i)},
\end {equation}
where $y(i)$ and $p_{x}$(i) are, respectively, the rapidity (calculated in the center of mass system) and
the momentum of the $i^{th}$ particle. The rapidity is defined as
\begin {equation}
Y(i)= \frac{1}{2}\ln\frac{{\vec{E}}(i)+{\vec{p}}_{z}(i)}
{{\vec{E}}(i)-{\vec{p}}_{z}(i)},
\end {equation}

where $\vec{E}(i)$ and $\vec{p_{z}}(i)$ are, respectively, the
energy and longitudinal momentum of the $i^{th}$ particle. In this
definition, all the rapidity bins are taken into account. It is worth mentioning that the E$_{bal}$
has the same value for all fragments types
\cite{west93,pak97,west98,cuss}. Further the apparatus
corrections and acceptance do not play any role in calculation of
the E$_{bal}$ \cite{ogli89,west93,cuss}.
\begin{figure}[!t]
\centering
 \vskip 1cm
\includegraphics[angle=0,width=12cm]{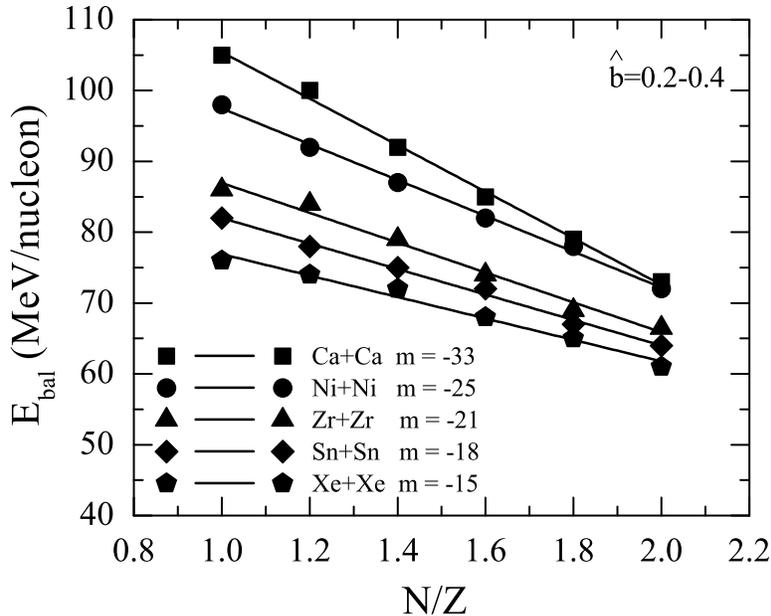}
 \vskip -0cm \caption{ N/Z dependence of E$_{bal}$ for various systems. Various symbols are explained
 in the text. Lines are linear fit. }\label{fig1}
\end{figure}

In Fig. 1, we display N/Z dependence of E$_{bal}$ for isotopes of
Ca+Ca (squares), Ni+Ni (circles), Zr+Zr (triangles), Sn+Sn
(diamonds), and Xe+Xe (pentagons). Lines are linear fit $\propto$
m$\ast$ N/Z.  From figure, we see that E$_{bal}$ decreases with
increase in N/Z for all the systems. The decrease in E$_{bal}$ is
linear with $|m|$ 33, 25, 21, 18, and 15 for Ca+Ca, Ni+Ni, Zr+Zr,
Sn+Sn and Xe+Xe series, respectively. Since we are having isotopes
of elements so Coulomb potential will be same throughout the given
isotopic series. Moreover, in recent study \cite{gaum310}, Sood
have shown that the effect of isospin dependence of cross section
is same for the higher N/Z range for Ca+Ca series, thus does not
affect the slope (m) of N/Z dependence of E$_{bal}$. This is
expected for other systems as well. For a given series as the N/Z
ratio increases, the neutron content also increases. Therefore,
the repulsion due to symmetry energy increases, hence the
E$_{bal}$ decreases with increase in N/Z.

\begin{figure}[!t]
\centering
 \vskip 1cm
\includegraphics[angle=0,width=12cm]{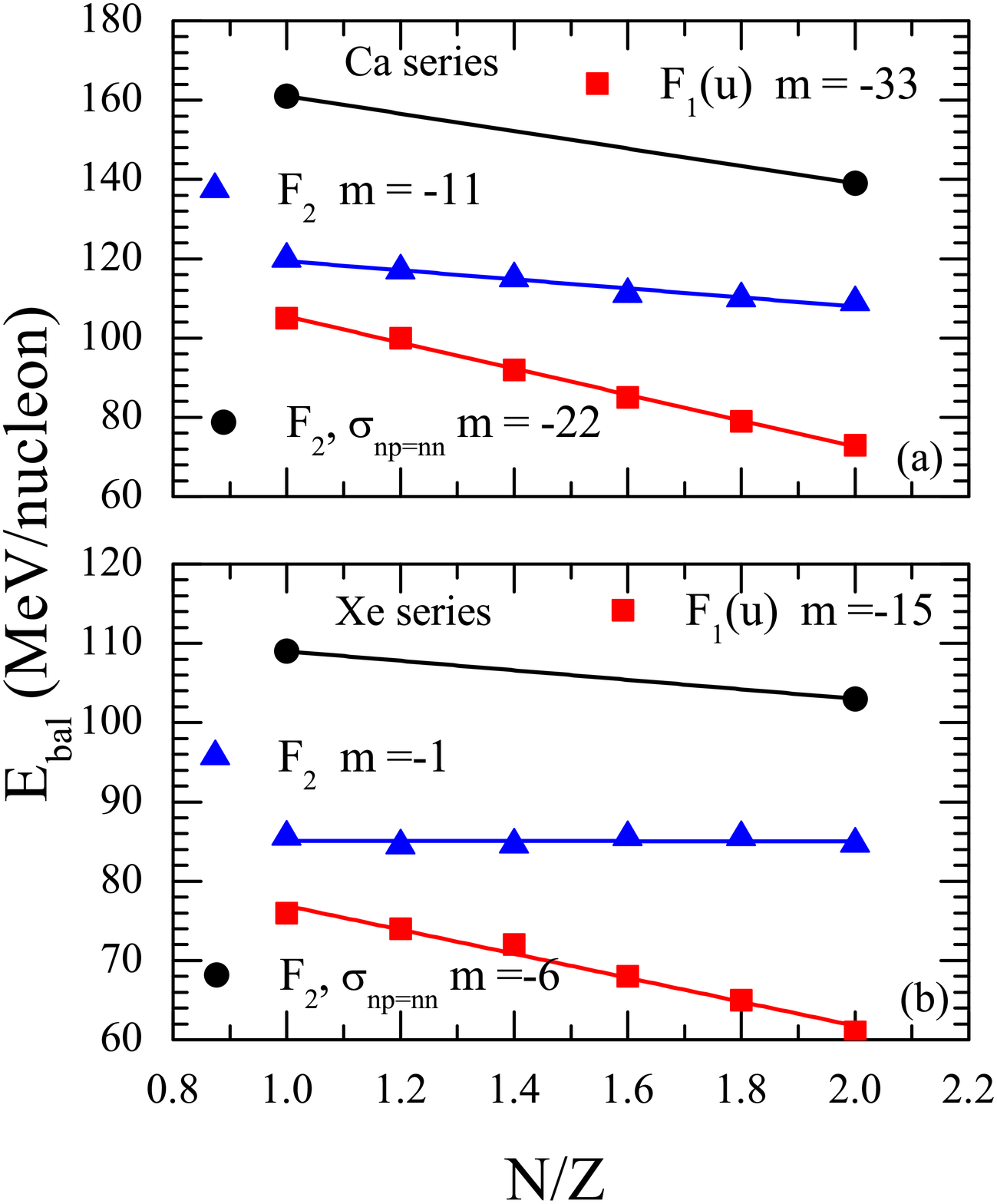}
 \vskip -0cm \caption{(Color online) N/Z dependence of E$_{bal}$ for Ca+Ca and Xe+Xe with E$_{sym}$ on and off. Various symbols are explained in text.
 Lines are linear fit.  }\label{fig2}
\end{figure}

 To see the effect of
symmetry energy on the N/Z dependence of E$_{bal}$ throughout the
mass range, we make the strength of potential part of symmetry energy zero
(represented by F$_{2}$, solid triangles) and calculate E$_{bal}$ for two extreme mass
systems Ca+Ca and Xe+Xe throughout the N/Z range. Results are
displayed in Fig. 2(a) and Fig. 2(b), respectively, for Ca+Ca and
Xe+Xe. Squares represent calculations with symmetry energy. We see that when we reduce the strength of
symmetry energy to zero, the slope of N/Z dependence of E$_{bal}$
decreases drastically for both Ca and Xe series. This indicates
that the effect of symmetry energy on the N/Z dependence of
E$_{bal}$  for a given isotopic series is of the same order
throughout the mass range. Interestingly, we find that E$_{bal}$
is same throughout the N/Z range for Xe series when we switch off
the symmetry energy. Whereas for Ca series there is small decrease
in E$_{bal}$ with N/Z. To further explore this point we make the
cross section isospin independent ($\sigma_{np}=\sigma_{nn}$) and
calculate E$_{bal}$ for two extreme N/Z = 1.0 and 2.0 for both
Ca+Ca and Xe+Xe reactions. The results are displayed by circles.
We see that the slope of N/Z dependence of E$_{bal}$ increases
when we switch off the symmetry energy and also take
$\sigma_{np}=\sigma_{nn}$ as compared to when we switch off
symmetry energy but cross section is isospin dependent. This is
because E$_{bal}$ decreases by large amount for N/Z = 1.0 for both
Ca+Ca and Xe+Xe reactions which indicates larger role of isospin
dependence of cross section for systems with N/Z = 1.0 as compared
to for systems with N/Z = 2.0 (see Ref. \cite{gaum310} also)
throughout the mass range. Therefore, the slope of N/Z dependence
of E$_{bal}$ decreases on inclusion of isospin dependence of cross
section. From Fig. 2 we also see the role of symmetry energy in systems with N/Z = 1 (compare triangle and square
at N/Z = 1) although symmetry energy contribution is not expected to be present in this case. To explore this, we
calculate for $^{40}$Ca+$^{40}$Ca, the transverse flow of particles having $\rho/\rho_{0}$ $\leq$ 1 (denoted as BIN 1) and
particles having $\rho/\rho_{0}$ $>$ 1 (denoted as BIN 2), respectively at all the time steps. The incident energy is taken to be 100 MeV/nucleon.
The results are displayed in Fig. 3.

\begin{figure}[!t] \centering
 \vskip -1cm
\includegraphics[width=12cm]{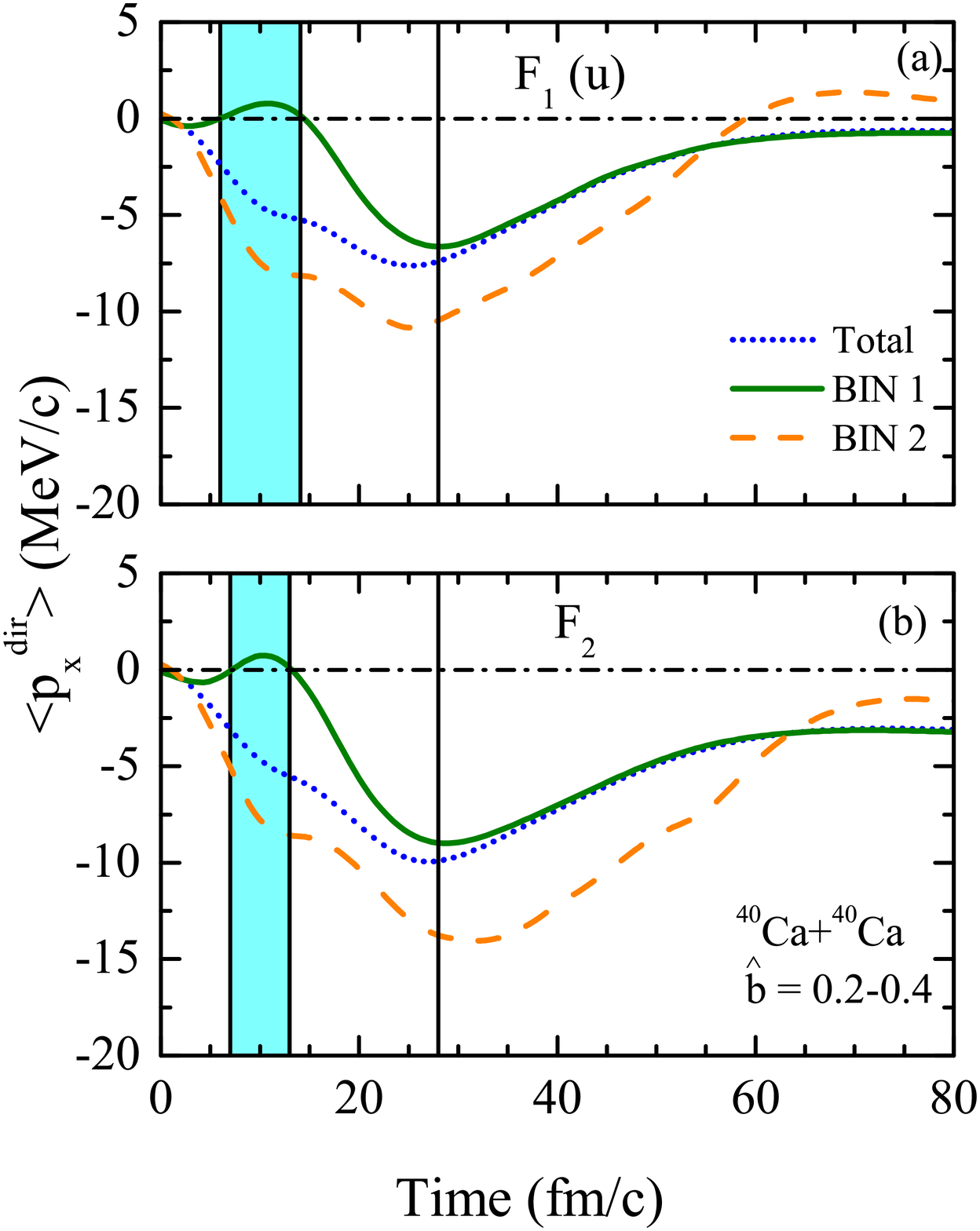}
\caption{(Color online) The time evolution of
$<p_{x}^{\textrm{dir}}>$ for $^{40}$Ca+$^{40}$Ca reaction for calculations with (upper panel)
and without (lower) symmetry energy at 100 Mev/nucleon. Lines are
explained in text.}\label{fig3}
\end{figure}

Solid (dashed)
lines represent the $<p_{x}^{dir}>$ of particles lying in BIN 1 (BIN 2). Dotted lines represent the total $<p_{x}^{dir}>$. Upper (lower)
panel is for calculations with (without) symmetry energy. We see
that the total $<p_{x}^{dir}>$ is sensitive to the symmetry energy even for symmetric systems like $^{40}$Ca+$^{40}$Ca with N/Z = 1.
This is because during the initial stages (between about 5-15 fm/c) the $<p_{x}^{dir}>$ due to particles in BIN 1 is positive. The duration for which
it remains positive is enhanced when we include the symmetry energy (compare shaded area in Fig. 3(a)and 3(b)). This
is because in the spectator region (where high rapidity particles lie), the repulsive (attractive) symmetry energy for neutrons (protons)
will accelerate the neutrons (protons) away (towards) the overlap zone. Inside the overlap zone the particles are stopped due to
collisions. Since these particles belong to same midrapidity region, their momenta (due to symmetry potential) will add up to the
zero thus nullifying the effect of symmetry potential for protons whereas the neutrons will end up in spectator rapidity region leading to a net
momentum due to the effect of symmetry energy. After about 15 fm/c, $<p_{x}^{dir}>$ of particles lying
in the BIN 1 becomes negative because these particles will now be attracted towards central dense zone. These particles will feel
the attractive mean field potential up to about 25 fm/c after which the high density phase is over. The decrease in total $<p_{x}^{dir}>$
due to attractive mean field potential (between 15-25 fm/c) is less when we include the symmetry potential in our
calculations (compare the slopes of dotted curves (total $<p_{x}^{dir}>$) in Fig. 3(a) and 3(b) between the right
edge of shaded area and vertical line).
This is because the neutrons due to the effect of symmetry energy lie in the spectator rapidity region with momenta away from the overlap zone.
The attractive mean field potential will have to decelerate those particles first, make them stop and then accelerate
the particles back towards the overlap zone. After about 30 fm/c,
 total $<p_{x}^{dir}>$ follows the $<p_{x}^{dir}>$ of particles lying in BIN 1 because of the expansion phase
 of the system. This explains the effect of symmetry energy in system with N/Z = 1. This effect is enhanced with increase
 in N/Z of the system \cite{gaum410}.

\begin{figure}[!t]
\centering
 \vskip 1cm
\includegraphics[angle=0,width=12cm]{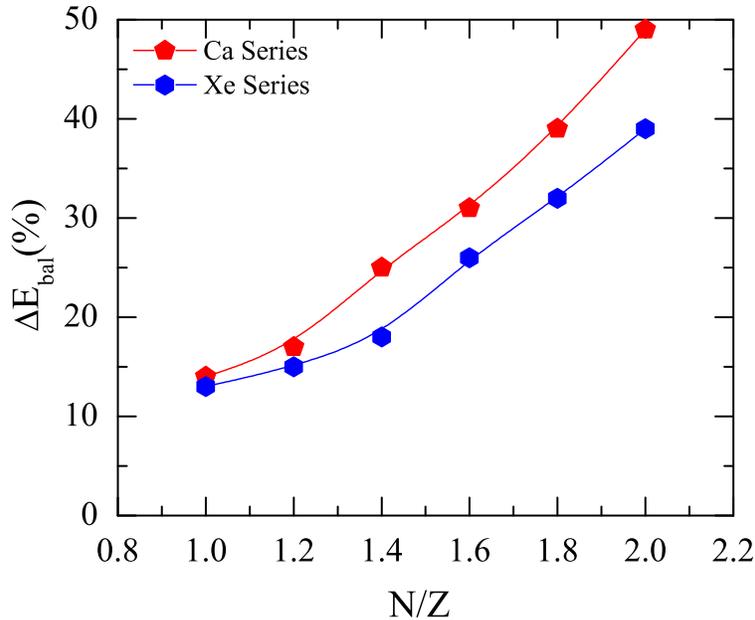}
 \vskip -0cm \caption{ N/Z dependence of $\Delta$E$_{bal}$($\%$) for Ca and Xe series. Various symbols are explained in text. Lines are only to guide the eye.}
 \label{fig4}
\end{figure}

\par
In Fig. 4, we display the percentage difference
($\Delta$E$_{bal}$($\%$)= $\frac{E_{bal}^{symm off}
-E_{bal}}{E_{bal}}$) between calculations without symmetry energy
and with symmetry energy as a function of N/Z for Ca (pentagons)
and Xe (hexagons) series. We see that the percentage difference
increases with increases in N/Z for both Ca and Xe series, which
shows that the effect of symmetry energy increases with increase
in N/Z. The increase is more sharp for Ca series as compared to
Xe, which indicates that with increase in N/Z the effect of
symmetry energy increases more sharply for Ca as compared to Xe
series. We also see that for N/Z = 1.0, the role of symmetry
energy is same throughout the mass range as far as E$_{bal}$ is concerned. Thus, the N/Z dependence
of $E_{bal}$ for the isotopic series of lighter systems is a
slightly better probe as compared to for heavier systems for
constraining the symmetry energy.
\par

\begin{figure}[!t]
\centering
 \vskip 1cm
\includegraphics[angle=0,width=12cm]{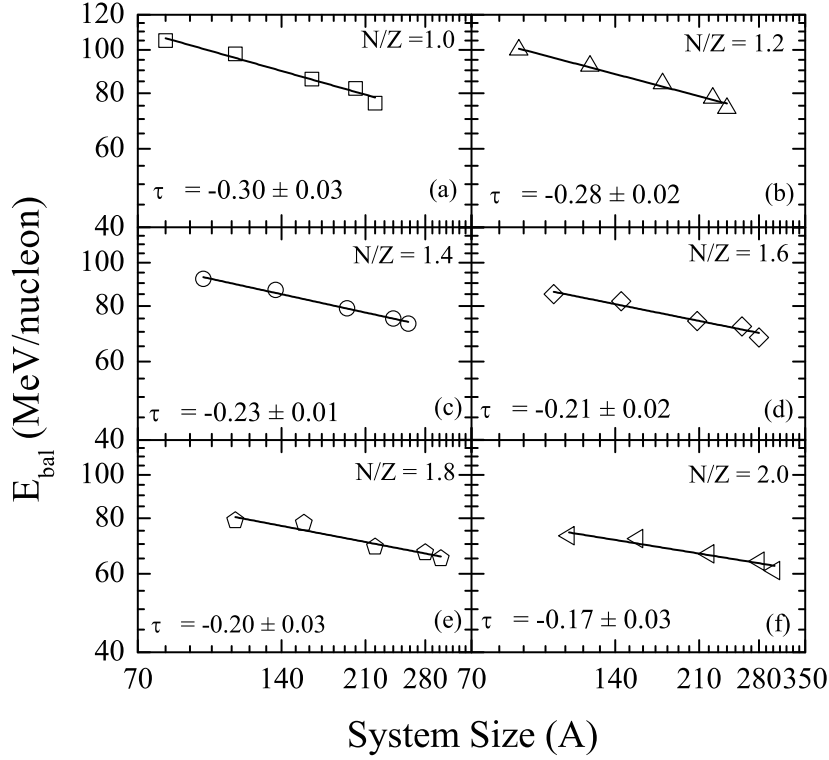}
 \vskip -0cm \caption{ System size dependence of E$_{bal}$ for various N/Z ratios. Lines are of power law fit ($\propto$ A$^{\tau}$).}\label{fig5}
\end{figure}

In Fig. 5, we display the system size dependence of E$_{bal}$
throughout the mass range. We find that for each N/Z, E$_{bal}$
follows a power law behavior ($\propto$ A$^{\tau}$) with power law
parameter $\tau$ = -0.30$\pm$0.03, -0.28$\pm$0.02, -0.27$\pm$0.01,
-0.21$\pm$0.02, and -0.20$\pm$0.03 for N/Z = 1.0 (open squares),
1.2 (triangles), 1.4 (circles), 1.6 (diamonds), 1.8 (open
pentagons), and 2.0 (left triangles), respectively. We find that
the value of $\tau$ decreases with increase in N/Z of the systems.
This is due to that fact that for higher N/Z ratio the effect of
symmetry energy is more in lighter masses (as discussed
previously), thus decreasing the E$_{bal}$ by larger magnitude on
inclusion symmetry energy in lighter masses which results in less
slope for higher N/Z ratio. Here we would like to stress that with increase
in N/Z for a given mass (compare mass of about 120 in Fig. 5(a) and 5(b)), the E$_{bal}$
decreases which is quite the opposite trend to the data \cite{pak97}. Since we have
taken the isotopes of a given element, so that the Coulomb is same for a given series. This
is because we have shown clearly in Ref. \cite{gaum10,gaum210} that Coulomb repulsion plays much more dominant role over symmetry
energy in isospin effects (if one considers isobars) throughout the mass range and colliding geometry.
\par

\section{Summary}
We have studied the N/Z dependence of balance energy (E$_{bal}$)
for isotopic series throughout the mass range. We found drastic
effect of symmetry energy on the N/Z dependence of E$_{bal}$
throughout the mass range. We also found that the N/Z dependence
of E$_{bal}$ for isotopic series of lighter system is slightly
more sensitive to symmetry energy as compared to that of heavier
systems. We have also studied the mass dependence of E$_{bal}$ for
the N/Z range from 1.0-2.0. We found that the mass dependence of
E$_{bal}$ varies with the N/Z ratio.

\par
This work is supported by a grant from Centre of Scientific and
Industrial Research (CSIR), Government of India and Indo-French
center vide project no-4101-A, New Delhi, India.



\end{document}